\documentclass[conference]{IEEEtran}
\IEEEoverridecommandlockouts
\usepackage{cancel}
\usepackage{amsmath,amssymb,amsfonts}
\usepackage{algorithmic}
\usepackage{graphicx}
\usepackage{textcomp}
\usepackage{xcolor}
\usepackage{url} 
\usepackage{flushend} 
\usepackage{subfig}
\usepackage{glossaries}
\def\BibTeX{{\rm B\kern-.05em{\sc i\kern-.025em b}\kern-.08em
    T\kern-.1667em\lower.7ex\hbox{E}\kern-.125emX}}

\begin{document}

\title{SmartSync: Cross-Blockchain Smart Contract Interaction and Synchronization
}

\author{\IEEEauthorblockN{Martin Westerkamp}
	\IEEEauthorblockA{\textit{Service-centric Networking} \\
		\textit{Technische Universit\"at Berlin}\\
		Berlin, Germany \\
		westerkamp@tu-berlin.de}
	\and
	\IEEEauthorblockN{Axel K\"upper}
	\IEEEauthorblockA{\textit{Service-centric Networking} \\
		\textit{Technische Universit\"at Berlin}\\
		Berlin, Germany \\
		axel.kuepper@tu-berlin.de}
}

\maketitle

\begin{abstract}
Cross-Blockchain communication has gained traction due to the increasing fragmentation of blockchain networks and scalability solutions such as side-chaining and sharding.
With SmartSync, we propose a novel concept for cross-blockchain smart contract interactions that creates client contracts on arbitrary blockchain networks supporting the same execution environment.
Client contracts mirror the logic and state of the original instance and enable seamless on-chain function executions providing recent states.
Synchronized contracts supply instant read-only function calls to other applications hosted on the target blockchain.
Hereby, current limitations in cross-chain communication are alleviated and new forms of contract interactions are enabled.
State updates are transmitted in a verifiable manner using Merkle proofs and do not require trusted intermediaries.
To permit lightweight synchronizations, we introduce transition confirmations that facilitate the application of verifiable state transitions without re-executing transactions of the source blockchain.
We prove the concept's soundness by providing a prototypical implementation that enables smart contract forks, state synchronizations, and on-chain validation on EVM-compatible blockchains.
Our evaluation demonstrates SmartSync's applicability for presented use cases providing access to recent states to third-party contracts on the target blockchain.
Execution costs scale sub-linearly with the number of value updates and depend on the depth and index of corresponding Merkle proofs.
\end{abstract}

\begin{IEEEkeywords}
Blockchain interoperability, Smart Contracts, Sidechains
\end{IEEEkeywords}

\newacronym{pow}{PoW}{Proof-of-Work}
\newacronym{pos}{PoS}{Proof-of-Stake}
\newacronym{spv}{SPV}{Simplified Payment Verification}
\newacronym{asic}{ASIC}{Application-Specific Integrated Circuit}
\newacronym{nipopow}{NIPoPoW}{Non-Interactive Proofs of Proof-of-Work}
\newacronym{evm}{EVM}{Ethereum Virtual Machine}
\newacronym{utxo}{UTXO}{unspent transaction output}
\newacronym{eoa}{EOA}{externally-owned account}
\newacronym{ens}{ENS}{Ethereum Name Service}
\newacronym{eip}{EIP}{Ethereum Improvement Proposal}
\newacronym{amm}{AMM}{Automated Market Maker}
\newacronym{xcmp}{XCMP}{Cross-Chain Messaging Protocol}

\section{Introduction}
During the last decade, blockchain technology has experienced rapid development resulting in a growing number of blockchain networks.
While novel forms of consensus algorithms~\cite{Kiayias2017Ouroboros,Buterin2020POS,Stewart2020Grandpa} and smart contract execution environments~\cite{khan2021trends} pose opportunities for a wide range of use cases, we observe an increasing fragmentation of applications across multiple blockchain networks.
This tendency is contradictory to the initial proposition of smart contracts to provide decentralized and highly available applications, as it constitutes a new form of lock-in effect towards the host blockchain.
Because the network is isolated, smart contracts are only interoperable with other smart contracts in the same network.
Current proposals to enable cross-chain function calls entail multiple transactions on the initiating blockchain and cannot cater use cases that require instant responses.
Furthermore, contracts need to be adjusted to support cross-blockchain calls, as specific message formats and fallback functions are required to process return values.

Due to the limitations of current cross-chain solutions, use cases that require up-to-date read-only access cannot be implemented efficiently.
For example, registries such as the \gls{ens} provide mappings between human-readable names and a plethora of attributes.
These attributes may include account addresses of a set of blockchains.
Hereby, a link is created between multiple wallets of a user on various blockchain networks that may be utilized for further processing or transfers.
By providing cross-chain access to such registries, users will be enabled to perform a single name registration on a blockchain of their choice and use it across multiple blockchain networks.

One of the most prominent application domains enabled by blockchain technology is Decentralized Fincance (DeFi), which provides financial products without involving intermediaries.
Most DeFi applications require trusted data, such as price feeds, that is typically delivered by oracles~\cite{liu2020defi}.
As an alternative to external oracles, \glspl{amm} like Uniswap have proven to provide reliable on-chain price estimations~\cite{uniswap2021oracles}.
Up until now, only smart contracts hosted on the same blockchain as the \gls{amm} could benefit from the provided data.
Allowing smart contract queries across blockchain networks will permit hosting applications on the best-suited blockchain while retrieving data feeds cross-chain.

We enable such use cases and remedy shortcomings of isolated smart contract states by developing a novel concept called \emph{SmartSync} for smart contract synchronization across blockchains.
Here, a secondary contract is created on the target blockchain that reflects the logic and current state of a smart contract on the source blockchain.
State updates are synchronized by regularly applying storage proofs that are verified based on a trusted storage root.
Our approach is agnostic to the storage root source, as long as it is trusted and available on-chain.
Thus, it is applicable to chain relays, notary schemes, sharding as proposed for Ethereum 2.0~\cite{Buterin2020POS} or related concepts such as Polkadot or Cosmos~\cite{Wood2016,Kwon2018}.
As synchronizations can be initiated within the same block of a function call, applications can rely on recent states.
While smart contract calls are currently limited to their host blockchain or shard, SmartSync provides synchronized replicas of remotely hosted smart contract instances.
These replicas provide all view functions of the original contract.
As a result, third-party contracts are enabled to execute the logic of a contract that was initially deployed on another blockchain network to retrieve verified information.
To prove the concept's soundness, we provide a prototypical implementation of SmartSync\footnote{https://github.com/disco-project/smart-sync} and publish it under an open source license.

Our contributions are as follows:
We propose a novel concept for synchronizing smart contracts across multiple blockchain networks and introduce a new form of state transition confirmation to facilitate trustless, lightweight state updates.
A prototypical implementation including off-chain client and respective smart contracts is provided and published.
The prototype is used to prove the concept's feasibility and evaluate implied operational costs.


\section{Preliminaries}
In this Section, we first introduce common terms that are used in the following.
Thereafter,  multiple approaches for providing state roots across blockchains are presented, as SmartSync utilizes derived Merkle proofs to enable trustless synchronizations.
Third, different approaches of cross-chain smart contract portability are categorized.
\subsection{Definitions}
We base our concept on the assumption that two blockchain instances \(A\) and \(B\) exist in parallel and support the same execution environment for smart contracts.
Information is only transmitted in one direction from \(A\) to \(B\).
Therefore, we also refer to \(A\) as \emph{source} and \(B\) as \emph{target} blockchain in the following.
To exemplify the applicability of our concept, we presume an Ethereum-like implementation of state storage for our model~\cite{Wood2014}.
Thus, the model incorporates an account-based model to represent the state rather than utilizing \glspl{utxo}~\cite{nakamoto2008}.

Each blockchain is composed of a sequence of block headers \(H\).
We denote \(H^A_t\) as the head of blockchain \(A\) at time \(t\) and \(H^A_{t+1}\) as the subsequent block header.
Block headers hold information proving their validity according to the applied consensus algorithm and a reference to the current state.
The global state is defined as the union of all account states.
Each account is identified by an address and owned by external users or smart contracts.
Every account holds a state that is composed of a set of key/value pairs \(f : K \rightarrow V\).

We define the state of contract \(C\) at time \(t\) and instantiated on blockchain \(A\) as \[S(C^A_t) = \{(k_1,v_1), (k_2,v_2), ... (k_n,v_n)\}.\]
The Merkle hash tree function \(m\) maps the entire state to a storage root \(R\), so that \(m: S(C^A_t)\rightarrow R({C^A_t})\). 
The Merkle proof \(\prod (R(C^A_t), (k,v))\) proves that \((k,v)\in S(C^A_t)\).
Equivalently, \(S(H^A_t)\) represents the global state and the respective Merkle hash tree function maps an account \(C^A_t\) to the global state root \(R(H^A_t)\), so that \(\prod (R(H^A_t), C^A_t)\) proves \(C^A_t \in S(H^A_t)\).
Note that \(S(C^A_t)\subseteq S(H^A_t)\).

\subsection{Cross-chain State Roots}
Account-based blockchains represent their global state in Merkle trees~\cite{Wood2014}.
The tree's root is referred to as \emph{state root} and enables proving the presence of a specific account state at a given time.
Providing a state root of a primary blockchain to a secondary blockchain permits accessing the primary blockchain's state.
In the following, we present three approaches for transferring state roots between blockchains.
\subsubsection{Chain Relays}
Chain relays reflect the consensus mechanism of a primary blockchain on a secondary blockchain~\cite{buterin2016interoperability}.
Block headers of a primary blockchain are submitted, validated and stored on a secondary blockchain if the validation was successful according the primary blockchain's consensus rules.
%
Once a block header is deemed final by the relay, it can be utilized to prove the inclusion of transactions, state or events within the primary blockchain on the secondary blockchain.
As block headers comprise corresponding Merkle roots, the submission of Merkle proofs enables inclusion proofs without requiring trusted intermediaries.
A plethora of chain relays exist that validate block headers within smart contracts~\cite{btcrelay}, off-chain programs using zkSNARKs~\cite{westerkamp2020zkRelay}, or by applying an optimistic approach~\cite{frauenthaler2020ethrelay}.
\subsubsection{Sharding}
The objective of blockchain sharding is to enhance throughput by distributing transactions and applications among multiple shards and processing them in parallel~\cite{Kokoris2018OmniLedger}.
While \gls{utxo} variants exist~\cite{Kokoris2018OmniLedger,Luu2016Sharding}, account-based implementations operate on a shared state root that is accessible through a hub blockchain~\cite{Wood2016,Kwon2018,Buterin2020POS}.
The shared state root subsumes the state of all connected shards and is relayed back from the hub to all shards.
As a result, cross-shard applications are enabled by processing states, transactions or events through Merkle proofs derived from the shared state root.
\subsubsection{Notary Schemes}
Notary schemes facilitate arbitrary information exchange between blockchains by introducing trusted notaries~\cite{buterin2016interoperability}.
As no complex validation is required, the scheme's implementation is comparatively simple.
For instance, transferred data may be trusted if it was signed by a predefined quorum of notaries.
\subsection{Smart Contract Portability}
The term \emph{smart contract portability} subsumes mechanisms that enable the migration of smart contracts between blockchain instances, including logic and state.
Smart contracts are typically hosted on a single blockchain and inherit its security and performance properties.
As a prerequisite for smart contract portability, source and target blockchains must support the same execution environment. 
We distinguish three types of smart contract portability: forks, moves, and the novel synchronization scheme presented in this paper.
\subsubsection{Smart Contract Forks}
\label{forks}
Similar to blockchain forks, smart contract forks describe the creation of a second instance that exists in parallel to the original instance and is based on a shared history~\cite{westerkamp2019portability}.
Smart contract forks are typically created on a second blockchain instance to overcome shortcomings of the primary host blockchain.
Reasons for creating smart contract forks on different blockchains are manifold:
\begin{itemize}
	\item Incentives of validators (or miners) of the host blockchain may not be aligned with those of smart contract users. While validators typically intend to receive high rewards, smart contracts seek for high throughput, short waiting times, and low fees.
	\item Different applications require different degrees of security. While a smart contract's target application may be aligned with the host blockchain's at the time of deployment, its security may decrease over time. 
	\item During a smart contract's life cycle, external requirements may change. 
	\item Over time, better suited blockchain networks may become available for a given use-case. 
\end{itemize}
To create a smart contract fork, the executing entity first retrieves the stored bytecode and state.
The state does not necessarily have to reflect the latest state, but it can also be retrieved from a preceding block, depending on the intention.
While the bytecode is easily retrievable using any synchronized full-node client, the state must be reconstructed, i.e. all key/value pairs must be retrieved.
Prior work suggested to reconstruct the state by retrieving all transactions ever sent to the target contract in order to observe each state transition~\cite{westerkamp2019portability}.
The current state is rebuild by applying all state transitions consecutively.
In case many transactions have modified the contract state, this process introduces significant overhead.
Today, node client implementations provide more elaborate traceability functions that permit retrieving all occupied keys of a smart contract.
In our synchronization concept, we use this mechanism to facilitate rapid contract forks during the initialization phase prior to synchronization.
\subsubsection{Smart Contracts Moves}
Fynn et al. have proposed a mechanism for moving smart contracts between ledgers~\cite{fynn2020move}.
In contrast to smart contract forks, where two or more instances of a contract exist in parallel, a smart contract move suspends the primary contract before enabling the secondary instance.
The authors propose dedicated instructions to the virtual machine to facilitate preconditional locking and unlocking of migrated contracts.
Beneficial to users is the decreased risk of dispersion across multiple instances.
On the downside, moving contracts is only feasible for authorized entities and users must adhere to the move decision.
Compared to this approach, smart contract forks favor the risk of  fragmentation over centralization.
\section{Smart Contract Synchronization}
The objective of smart contract synchronization is to provide instant \emph{read-only} access to a contract that was initially deployed to another blockchain.
Instant access is characterized as on-chain function invocation that terminates in a single transaction.
A secondary contract is created on the target blockchain that mirrors the original source contract.
Hereby, the smart contract's functionality and state becomes available to other smart contracts on the secondary blockchain for retrieving information, e.g. through getter methods, while changing its state is prevented.
Consequently, smart contracts hosted on a remote blockchain can be used within a single transaction, as the state migration process is executed previously in a verifiable manner.
Furthermore, complex queries can be achieved on-chain following the original contract logic.

As a prerequisite, a smart contract fork is created to migrate logic and current state of the target contract onto a secondary blockchain.
In contrast to smart contract forks, however, only those functions are callable that do not modify the state.
As the primary contract exists in parallel and remains callable, the states of primary and secondary contract may diverge over time.
We present a synchronization mechanism that ensures only valid state updates are applied.
\subsection{Synchronization Conditions}
\label{sync_conditions}
As the contract state may divert from a forked instance, a mechanism for synchronizing the state over time is required.
Given the decentralized nature of blockchain applications, we derive the following conditions that must be met to guarantee valid state updates:
\begin{itemize}
	\item\textbf{Equality.} The secondary contract's state must reflect a state of the primary that was valid at a given point in time according to the source blockchain's consensus rules. 
	\item\textbf{Completeness.} The entire state must be captured, including new, updated and deleted variables.
	\item\textbf{Succession.} Each synchronization update must reflect a state that succeeded the previous state.
	\item\textbf{Trustlessness.} The execution must not depend on any trusted intermediary.
\end{itemize}
Frequent synchronizations are required to provide valid states over time.
Thus, the depicted conditions are expected to be met at the time of synchronization.
The frequency must be defined by the referencing contract, as different applications may have distinct needs in terms of timeliness.
\subsection{Initialization Process}
\label{initialization}
Before a smart contract can be synchronized, its state and logic must be duplicated via a smart contract fork.
While prior work relied on external validation~\cite{westerkamp2019portability} in the process, we introduce a verifiable migration scheme that utilizes a trusted state root.
Before the source contract's state can be applied on the target blockchain, its bytecode and state must be retrieved either by replaying past transactions or using a blockchain client's extended database access.

\textbf{Contract separation.} First, the original contract is split into dedicated logic and proxy contracts.
The proxy contract maintains the contract state and delegates all function calls to the logic contract, which in turn executes function logic based on the proxy contract state.
This pattern enables the implementation of migration and synchronization logic within the proxy contract while keeping the source contract's logic separated.
Hereby, the contract logic's validity is maintained, as it is not modified.
The proxy contract supplies setter methods that enable distributing the initialization over multiple transactions.
After the entire state was replicated, the function cannot be invoked anymore, ensuring the state's integrity.
For further details on how to prevent future invocations without polluting the proxy contract's state, we refer to~\cite{westerkamp2019portability}.

\textbf{Verification.}
While external verification of the resulting state was required previously, we introduce a mechanism that ensures validity on-chain.
For this purpose, a dedicated SmartSync contract is introduced on the target blockchain that traces and verifies all migrated smart contracts from a specific source blockchain.
After successfully migrating a contract, a transaction is submitted to the SmartSync contract indicating the migration is finished.
To prove its correctness, two Merkle proofs are submitted:  \(\prod (R(H^A_t), C^A_t)\) and \(\prod (R(H^B_{t+x}), C^B_{t+x})\), where \(x\geq 1\).
They both depict account proofs indicating the existence of a specific account tuple on the respective blockchain.
The first proof guarantees the submission of a correct contract state and is verified using the source blockchain's state root.
The second proof confirms the correct application of the contract's source state on blockchain $B$.
We assume header \(H^B_{t+x}\) is available to the contract as an historic state.
For instance, Ethereum permits accessing header hashes of the 256 most recent blocks~\cite{Wood2014}.
In case the target ledger does not provide the respective block hash to smart contracts, the contract state's Merkle root can be computed within the SmartSync contract during submission.
%
\subsection{Synchronization Process}
The objective of smart contract synchronization is to provide a secondary instance of a smart contract on a remote blockchain.
The contract state is synchronized regularly, providing access to its current state to arbitrary contracts on the secondary blockchain.
The synchronization process is unidirectional, thus, the primary contract remains unaffected of derived synchronized contracts.
Furthermore, the secondary contract's state is only modifiable by proving the state was obtained from the primary contract.
As a prerequisite, we assume a smart contract fork is created and verified as depicted in Section \ref{initialization}.
To capture all conditions and challenges for smart contract synchronization, we first introduce a na\"ive approach and elaborate our concept based on it in the following.
\begin{figure}[!t]
	\centering
	\includegraphics[width=.99\linewidth]{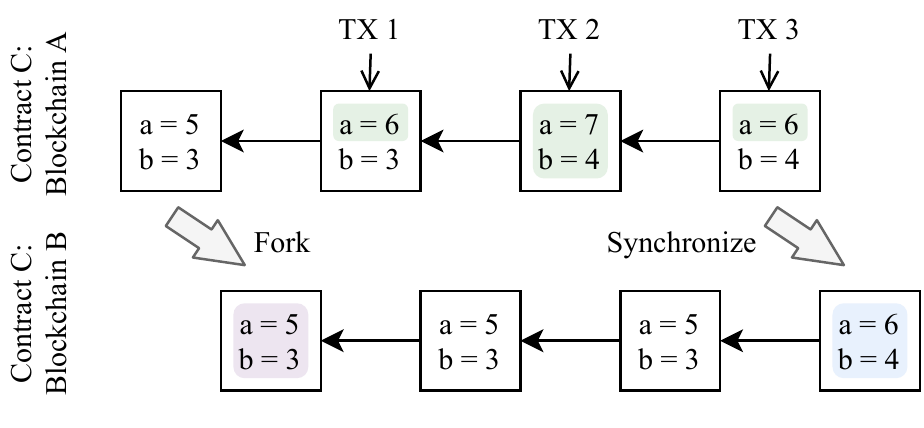}
	\caption{Simplified  contract state capturing the initial migration and synchronization of a diverted state. Green highlighting indicates transaction triggered value changes, purple illustrates the initial fork state and blue marks synchronized values.}
	\label{fig:overview_sync}
\end{figure}
\subsubsection{Na\"ive approach}
The state of smart contracts is the result of executed contract logic that is triggered by transactions.
Transactions are either initiated externally using an \gls{eoa} or internally by other smart contracts~\cite{Wood2014}.
As the objective of smart contract synchronization is to replicate the primary contract's state and the state is derived from transactions, it is intuitive to record all past transactions sent to the contract on the primary blockchain and apply them on the secondary blockchain.
For instance, Figure~\ref{fig:overview_sync} illustrates a contract $C^A$ that is forked and deployed to blockchain $B$.
In the following, three transactions are sent to $C^A$ modifying its state.
To reflect the resulting state on $C^B$, all three transactions would have to be executed sequentially on blockchain $B$.
The synchronization contract first verifies the inclusion of each transaction on $A$ using a Merkle-proof attesting its inclusion in a block that is provided by the relay contract.
Thereafter, the correct sequence is confirmed, i.e. no transaction may be skipped and the submission must adhere to the correct order.
By re-executing all transactions, the same state will be reached as on the source contract.
However, multiple challenges exist following such an approach.

Transactions cannot be re-executed natively, as the account matching the signing key pair either does not exist on the target blockchain or is likely to be in a distinct state, compared to the source blockchain.
Furthermore,  corresponding accounts must hold sufficient funds to compensate for transaction fees and a linked nonce is incremented every time a transaction is sent.
Thus, all accounts that have sent transactions to contract $C$ on blockchain $A$ must hold sufficient funds and be in an equivalent state.
In addition, transactions are targeted to a specific address necessitating equivalent addresses on both blockchains.
As these requirements can never be guaranteed, native execution of source transactions are not applicable.

Alternatively, transactions could be executed in an emulated environment.
For instance, in case of Ethereum, the \gls{evm} could be implemented within a smart contract that executes every transaction decoupled from the sender's account state and contract address on the target blockchain.
However, such emulation will come at the cost of increased complexity and thus execution costs.
Despite the inherent additional fees for synchronization, it cannot be guaranteed that all transactions can be replicated, as the limit of execution steps within a block may be exceeded.
As a result, an attacker could create a transaction executing complex logic to prevent its replication on the target ledger, effectively stalling contract synchronization.

Replaying every single transaction also incurs overhead with regards to storage operations.
Figure \ref{fig:overview_sync} presents an example where variable \texttt{a} is first changed from six to seven before being changed back to six.
While the result is equivalent, these operations cannot be subsumed and must be executed sequentially.
Moreover, to prove the inclusion of every transaction, respective storage roots must be provided by the relay contract.
As a result, optimized concepts that reproduce only a subset of historic storage roots, such as zkRelay~\cite{westerkamp2020zkRelay}, cannot be utilized, inducing overhead during synchronization.
\subsubsection{Verifiable state updates}
\begin{figure*}[!t]
	\centering
	\includegraphics[width=0.95\textwidth]{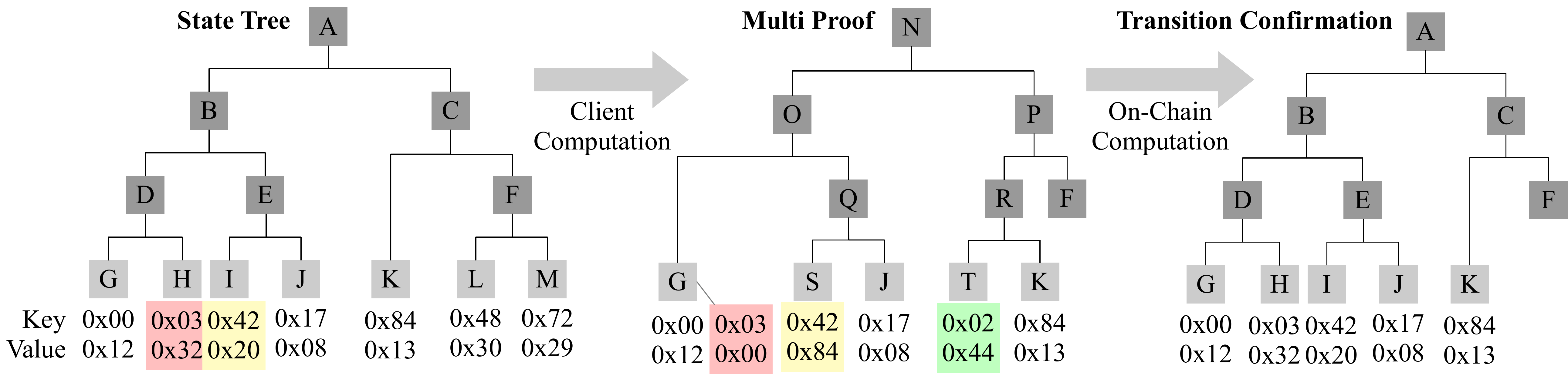}
	\caption{Left: Merkle tree of an account state derived from source chain. Center: Multi proof of an updated state: 0x03 is deleted (red), 0x42 is modified (yellow) and 0x02 is added (green). Right: Transition Confirmation calculated on the target chain.}
	\label{fig:trie}
\end{figure*}

We present a method that utilizes state proofs for synchronizing contract states across blockchains to solve the shortcomings of simple transaction replication.
Instead of proving the inclusion of transactions sent to the source contract, the  resulting state is used for updates.
To prove the presence of a particular key-value pair in the contract state, two Merkle proofs are required.
First, the account proof \(\prod (R(H^A_t), C^A_t)\) documents the inclusion of an account tuple within a block header $H$, where $R(H^A_t)$ is the global state root at time $t$.
Second, storage proofs notarize the inclusion of a specific key-value pair \(\prod (R(C^A_t), (k,v))\) in the account state, where $R(C^A_t)$ is is the account state root at time $t$.
Applying these proofs on the target ledger permits proving the addition, change or deletion of single key-value pairs.
Proving the result of executed transactions rather than replaying state transitions results in more efficient state updates in case of multiple write operations at a single key location.
For instance, Figure~\ref{fig:overview_sync} depicts multiple updates of multiple values.
Utilizing storage proofs permits to skip all intermediary changes and apply the verified final result.

The application of single key-value proofs meets all conditions defined in Section~\ref{sync_conditions} but \emph{Completeness}.
\emph{Equality} is met as the combination of account and storage proof ensures the presented key-value pair existed on blockchain $A$ at some time.
Only updates newer than the previous update are accepted.
\emph{Succession} and \emph{Trustlessness} are established via the utilized state root.
Furthermore, anyone can submit account and storage proofs, rendering the entire process verifiable.
\emph{Completeness}, however, is not assured, as storage proofs are independent from each other.
An attacker could submit only a subset of state updates to induce an inconsistent state.
While it would be impossible to add variables that are not part of the source contract, delete variables that haven't been deleted or modify values that haven't been modified, concealing a subset of state updates poses a serious vulnerability.
Figure~\ref{fig:trie} depicts a state transition $S(C^A_t)\rightarrow S(C^A_{t+x})$, where $x>0$ and includes three updates.
To fully capture the state transition, all three updates must be applied via dedicated storage proofs.
However, this would require trust in the executing entity and therefore violate the \emph{Trustlessness} condition.
Moreover, submitting a single storage proof for each modification creates considerable overhead, as each proof must be verified independently.
To overcome both issues, we introduce transition confirmations based on multi proofs.
\subsubsection{Multi proofs}
Multi proofs describe a mechanism for proving the inclusion of multiple values using a single proof and have been proposed to overcome inefficiencies when querying multiple values~\cite{Mizrahi2021MultiProof}.
While common Merkle proofs reflect a path within the respective Merkle tree, multi proofs share nodes of paths and thus depict sub-trees with substituting hash values where the tree was pruned.
As a result, a set of key-value pairs can be included in a proof \(\prod (R(C^A_t), \{(k_1,v_1), (k_2,v_2), ...(k_n,v_n)\})\), indicating that \(\{(k_1,v_1), (k_2,v_2), ... (k_n,v_n)\}\in S(C^A_t)\).
For instance, to prove the inclusion of keys $0x00$ and $0x03$ in Figure~\ref{fig:trie}, only the leaf values and two branch hashes of layers one and two of the tree are required.
The efficiency of multi proofs depends on the index of the target leaf values.
The more branches are shared, the more efficient a multi proof becomes in comparison to single proofs.

While multi proofs increase efficiency, they do not solve the threat of withholding parts of the resulting state.
Attackers could submit valid multi proofs including only a subset of key-value pairs involved in the original state transition.
Updating the secondary contract state by applying a multi proof could thus result in an inconsistent state.
\subsubsection{Transition confirmations}
The objective of transition confirmations is to guarantee \emph{Completeness}.
Thus, the entire state transition result must be captured, including all create, update and delete operations.
Transition confirmations are not submitted externally, but computed on-chain based on a multi proof.
As the state of the source contract is not accessible from the target ledger\footnote{Submitting Merkle proofs for each primary key-value pair would be possible but require considerable overhead that is avoidable by accessing the currently synchronized state on the target ledger.}, we utilize the fact that the latest target contract state $S(C^B_{t+x})$ equals the primary state $S(C^A_{t})$ of original state transition $S(C^A_t)\rightarrow S(C^A_{t+y})$, where $y>x>0$.
The submitted multi proof \(\prod (R(C^A_{t+y}), \{(k_1,v_1), (k_2,v_2), ... (k_n,v_n)\})\) indicates inclusion of respective key-value pairs given the Merkle root, but fails to create a link to the previous state $S(C^A_t)$.
Transition confirmations create such a link by replacing all values of the multi proof's key-value set with the original values of the state transition.
The original values are available during the computation, as they constitute the target contract's current state $S(C^A_t) = S(C^B_{t+x})$.
Therefore, they are retrievable by reading from the respective key location on the ledger.
While the multi proof's values are substituted, the actual paths remain equivalent.
Hereby, it is ensured that the entire state was captured, as withheld state updates would be reflected in changed paths.
The computed transition confirmation depicts a valid multi proof of the initial state $S(C^B_{t+x})$ if all changed values of the state transitions are reflected in the submitted multi proof:
\begin{gather*}
	\prod (R(C^A_{t+y}), \{(k_1,v^{t+y}_1), (k_2,v^{t+y}_2), ... (k_n,v^{t+y}_n)\})\rightarrow \\
	\prod (R(C^B_{t+x}), \{(k_1,v^{t+x}_1), (k_2,v^{t+x}_2), ... (k_n,v^{t+x}_n)\})
\end{gather*}
Thus, the transition confirmation is validated equivalently to a multi proof and we expect the computed Merkle root to equal the previous state's Merkle root if all changes were captured and different otherwise, as illustrated in Figure \ref{fig:trie}.
The rationale is that a multi proof is a pruned version of the original Merkle tree and includes those key-value pairs that have been modified during a state transition.
Withholding a subset of updates will be disclosed by this kind of transition check, as the withheld set is captured in a hashed branch:
\begin{gather*}
\prod (R(C^A_{t+y}), \{(k_1,v^{t+y}_1), \cancelto{\emptyset}{(k_2,v^{t+y}_2)}, ... (k_n,v^{t+y}_n)\})\nrightarrow \\
\prod (R(C^B_{t+x}), \{(k_1,v^{t+x}_1), ... (k_n,v^{t+x}_n)\})
\end{gather*}
When computing the transition confirmation's Merkle root, it becomes evident that the result is different from the original root in case the submitted multi proof is incomplete.

While the sole validation of a multi proof could not capture the state transition depicted in Figure \ref{fig:trie}, the computed transition confirmation ensures correct state updates.
For instance, let us assume an attacker submits a multi proof including state updates at keys $0x03$ and $0x24$ but withholding the value addition at key $0x02$.
To create a transition confirmation, the values at key locations $0x03$ and $0x42$ are read from the target ledger and inserted in the multi proof to build the transition confirmation.
As Node P remains unaffected, the tree root would be calculated from nodes $B$ and $P$ instead of $B$ and $C$, resulting in a hash different to $A$.
Thus, it becomes evident that an incomplete multi proof was passed and the state update is aborted.
\subsubsection{Workflow}
\begin{figure}[!t]
	\centering
	\includegraphics[width=.98\linewidth]{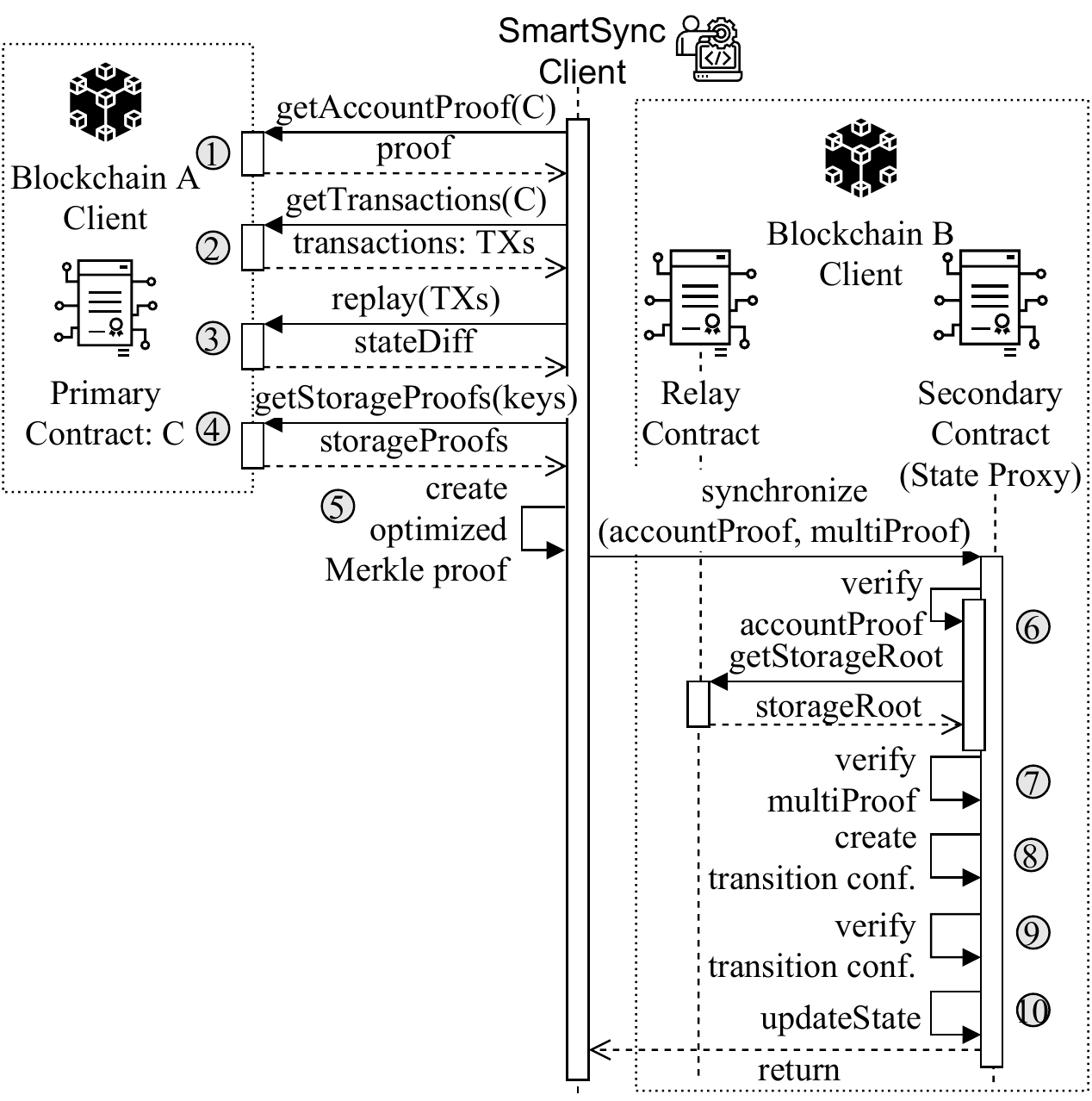}
	\caption{Synchronization workflow of SmartSync.
	}
	\label{fig:sequence}
\end{figure}

The combination of account proof, multi proof and on-chain computed transition confirmation enables verifiable state replication of smart contracts across multiple blockchain instances.
In the following, we depict the required steps to synchronize state updates from a primary contract to a secondary client contract hosted on a remote blockchain.
The entire workflow is illustrated in Figure \ref{fig:sequence}.

The process is initiated by any entity motivated to update an expired state on the target blockchain.
Motivation may be intrinsic as a depending smart contract of interest requires a recent state or it may be  triggered externally through a monetary incentive system.
As external incentive layers exceed the focus of this work, we refer to \cite{btcrelay,teutsch2019scalable,frauenthaler2020ethrelay} for further reading and leave its application on the presented use case for future work.
The external entity runs a SmartSync client having access to clients of blockchains A and B.

After retrieving the account proof from the source client, all transactions sent to the smart contract since the last update are retrieved and replayed locally (Steps 1-3).
Replaying respective transactions results in a set of key-value pairs representing the consequential state update.
Thereafter, a storage proof is retrieved from the source node for each key-value pair (Step~4) and combined to a multi proof (Step~5).
Both, account and multi proof are submitted to the proxy contract on the target blockchain in Step~6.
The account proof verification takes place within the proxy contract and presumes a respective block header to be available within the relay contract.
In case the block header is unavailable, the relay contract must be updated accordingly before starting the synchronization process.
After verifying the multi proof, it is used for creating the transition confirmation that ensures all state updates are included in the multi proof (Steps 7-8).
While the multi proof utilizes the Merkle root stored in the passed account proof, i.e. the new storage root, the transition confirmation is based on the current storage root that is stored within the smart contract and updated for each synchronization.
If the transition confirmation is deemed valid, the state updates included in the multi proof are applied to the proxy state (Steps 9-10).

Utilizing the resulting storage state of the source contract permits subsuming multiple transactions, reducing storage and execution costs.
In case the state update is too large to be executed within a single transaction, a previous state subsuming less transactions must be utilized first.

\section{Evaluation}
To evaluate SmartSync, a prototypical implementation of the concept was created that permits forking and synchronizing smart contracts between blockchains supporting the \gls{evm}.
The implementation is published under an open source license.
The prototype provides a client including command line interface to retrieve states and proofs from the source blockchain, create corresponding multi proofs and submit them to the target blockchain.
Furthermore, a proxy contract is supplied that retrieves, validates and applies state updates, delegates read-only function calls to the logic contract and prevents write operations other than synchronizations.
To prevent the delegation of function calls that would modify the state, the execution context must be static.
When retrieving a function call, the proxy contract's fallback function is executed, as it is the default operation in case the function signature is unknown.
The fallback function confirms the execution context is static and delegates the call to the logic contract.
Hereby, the executed logic operates on the state of the proxy contract that is a replica of the source contract.

\begin{figure*}[!t]
\centering
\subfloat[Updating a single value at different storage locations in the storage tree\label{fig:single_value}]{
	\includegraphics[width=.38\linewidth]{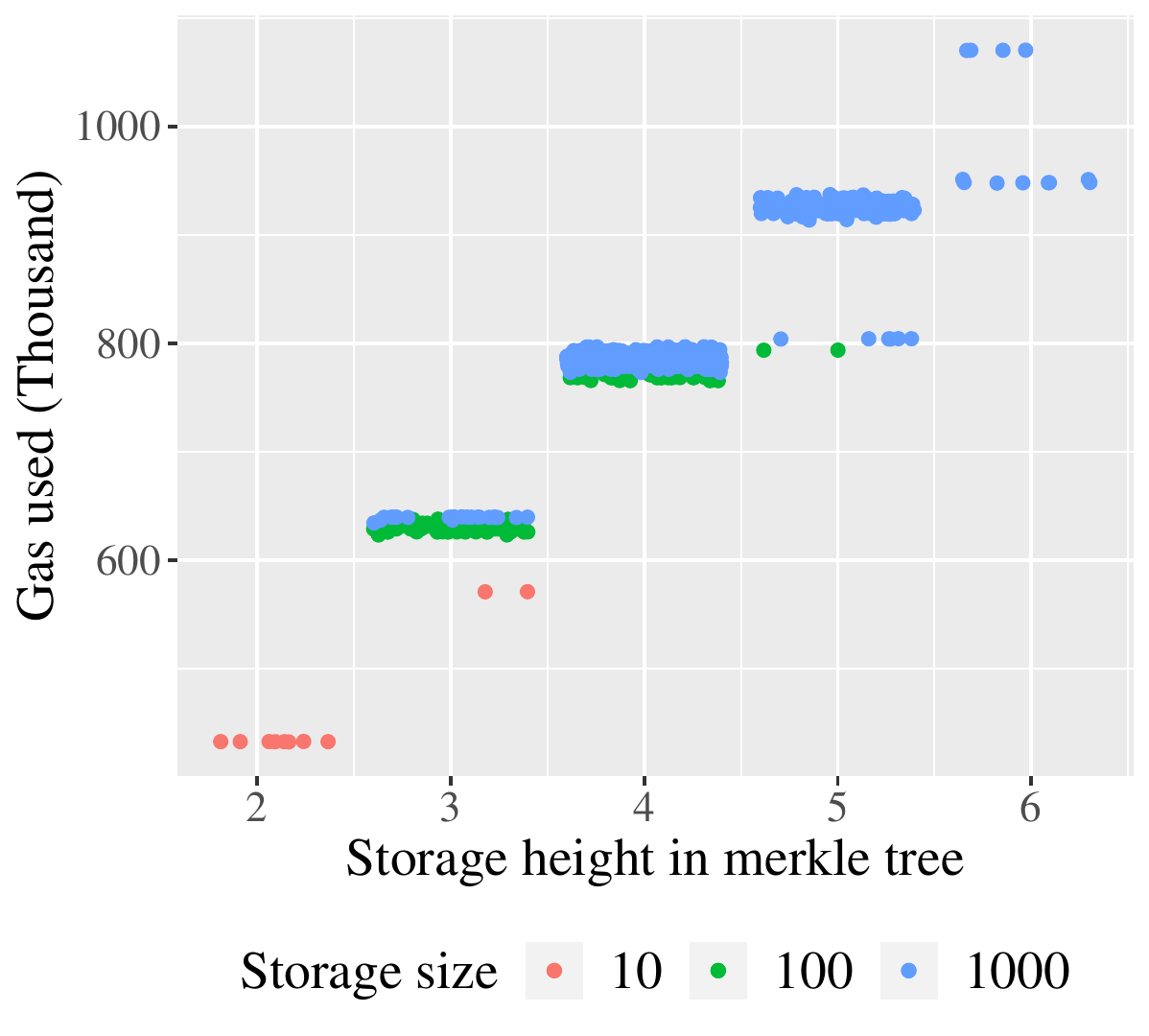}
}
\hfil
\subfloat[Updating multiple values using a single multi proof\label{fig:multi_values}]{
	\includegraphics[width=.38\linewidth]{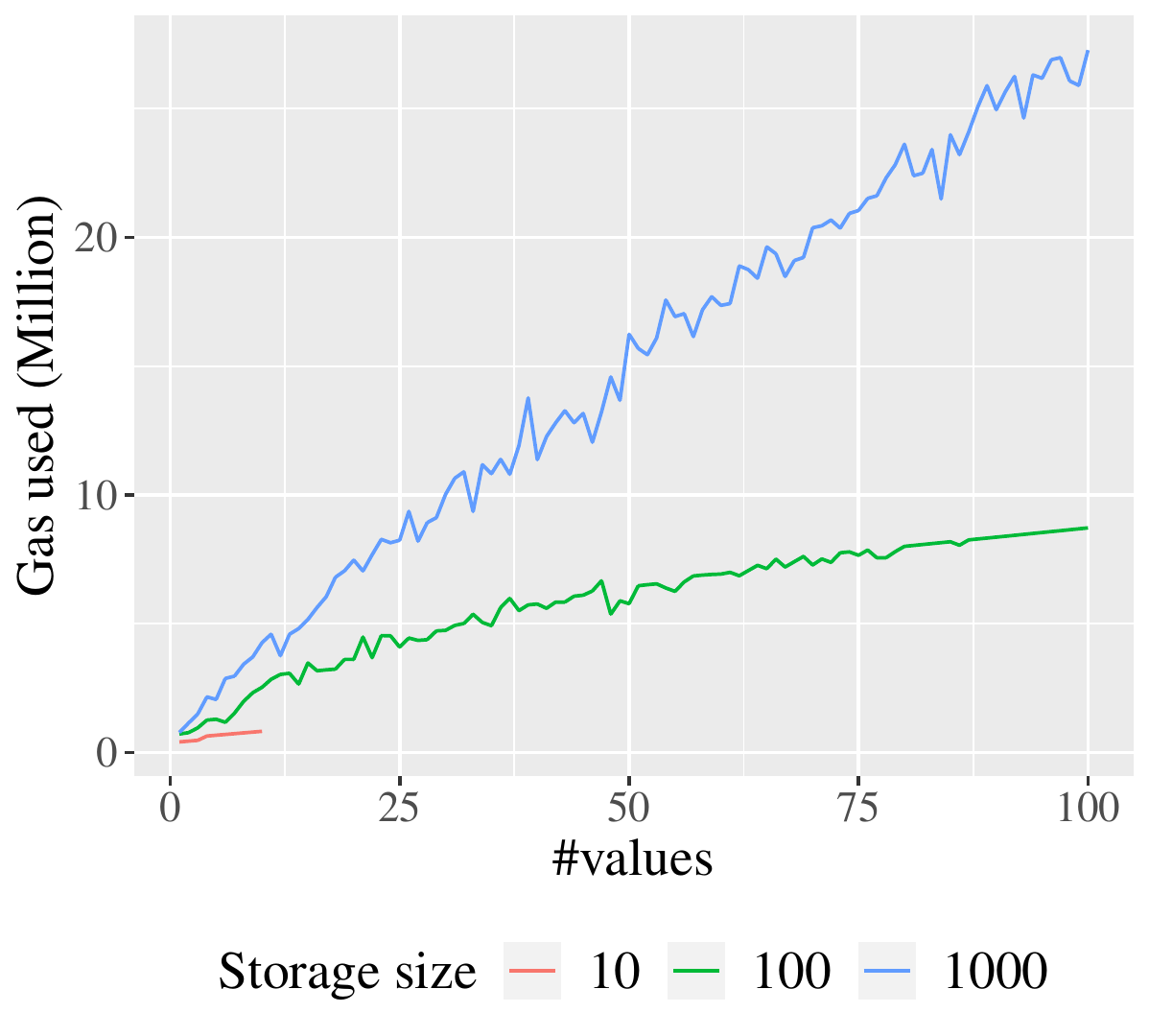}
}
\caption{Gas costs of synchronization processes including different update set sizes.}
\label{fig:evaluation}
\end{figure*}

The application of smart contract synchronization entails execution costs on the target blockchain.
As SmartSync is agnostic to the state root's source, we neglect the evaluation of corresponding application costs.
Ethereum measures execution costs by assigning a fee called \emph{gas} to every operation of the \gls{evm}~\cite{Wood2014}.
To determine the gas costs of synchronization operations, multiple contracts are deployed and synchronized.
Each contract holds a mapping of different size, resulting in storage tries of variable depth.
First, we analyze the costs of updating a single value depending on its index in the Merkle tree.
As indicated in Figure~\ref{fig:single_value}, the costs increase with the value's storage height within the tree, as the corresponding Merkle proof requires more nodes to be submitted.
Each additional node results in increasing costs for parameter size and proof validation.
Large contract storage sizes lead to deeper Merkle trees and thus drive synchronization costs.
Furthermore, execution costs depend on the node types included in the Merkle proof.
Ethereum utilizes modified Merkle Patricia trees to store values~\cite{Wood2014}.
We observe that proofs containing more extension nodes are cheaper than those containing branch nodes, as depicted in Figure~\ref{fig:single_value} for heights 3, 5 and 6.
As branches contain up to 17 elements, more bytes must be passed compared to extension nodes holding a two items.

Second, we evaluate the gas costs when updating multiple values in a single synchronization transaction.
As for single values, updates incur less gas costs for smaller storage sizes, as illustrated in Figure~\ref{fig:multi_values}.
Synchronization costs per value decrease with an increasing number of values subsumed in a single transaction.
Thus, gas costs increase sub-linearly.
As multi proofs share nodes of the original Merkle tree, fewer nodes have to be submitted and validated, the more nodes are shared.
The effect can be observed in Figure~\ref{fig:multi_values}, indicated by jitter in gas costs for each storage size.

Lastly, we measure the latency of synchronization updates including the retrieval of state proofs, computation of multi-proofs and execution on the target blockchain.
The benchmark was executed on a machine equipped with an Intel Core i7-6700HQ 266 CPU at  2,60\,GHz and 16\,GB memory.
Source and target blockchains are private Ethereum networks performing instant mining.
Our results show that updating a single value takes between 91\,ms and 475\,ms for updating a single value, depending on the Merkle tree's depth.
An update transaction that includes multiple values requires 119\,ms for 10 values and grows linearly to 1,879\,ms for 1,000 values.
As the involved latency remains well within the block mining times of most blockchain networks, the concept is suitable for applications requiring timely updates.
\section{Related Work}
Most cross-chain protocols target the exchange or transfer of tokens across blockchain networks~\cite{zamyatin2019xclaim,Gazi2019,Herlihy2018AtomicSwaps,Borkowski2019Dextt,Herlihy2021cross}.
As SmartSync captures elaborate smart contract functionality going beyond token handling, we present concepts proposing cross-chain smart contract interaction in the following.

HyperService is a concept for enabling cross-chain smart contract invocations~\cite{Liu2019HyperService}.
The approach includes a domain-specific language that generates decentralized applications utilizing multiple interacting smart contracts which are deployed to separate blockchain networks.
A state model is created to track the global application state on a dedicated blockchain for cross-chain communication.
Thus, a hub blockchain orchestrates cross-chain smart contract interaction.
In contrast, SmartSync enables instant smart contract queries after synchronization and does not implement any abstraction to enable access to decentralized applications.

Nissl et al. have proposed a concept that also enables cross-chain function calls, but does not implement an abstraction layer and does not require a mediating blockchain to orchestrate interaction~\cite{nissl2021crosschain}.
Instead, smart contract invocations are executed by intermediaries and checked by validators who finalize transactions.
In case misbehaviour is obeserved, deposited funds get reduced.
While the concept permits cross-chain function calls, multiple contract invocations and waiting times are required.
Furthermore, the approach is only secure under economic assumtions, while SmartSync derives its security from Merkle proofs and a trusted state root.
Chainbridge\footnote{https://chainbridge.chainsafe.io/} constitutes a similar protocol but relies on a trusted federation rather than economic safeguarding.
In comparison, SmartSync permits instant read-only function calls of synchronized contracts but cannot invoke functions across blockchains.

The \gls{xcmp}\footnote{https://research.web3.foundation/en/latest/polkadot/XCMP/} enables sending messages between shards in the Polkadot ecosystem~\cite{Wood2016}.
As shard communication is implemented via the relay chain mechanism, delays are low and its setup is comparable to SmartSync.
However, it does not focus on smart contract replication across shards but facilitates communication by providing dedicated message queues mediated by bilateral channels between shards.
Instant function calls are not possible and smart contracts must implement the message mechanism.






\section{Discussion}
Different use cases have specific requirements on the refresh period of depending smart contracts.
For instance, DeFi applications typically rely on up-to-date data, while registries such as the \gls{ens} do not change as regularly.
SmartSync can cater for both use cases given the storage root is updated in time.
Chain relay and target contract can be updated in the same block as the calling contract execution to provide the most recent state.
In case SmartSync is applied to provide accessibility across shards, the utilization of synchronized contracts may behave equivalently to locally hosted contracts.
This is the case if the calling contract requires a state of the latest created block and the block creation time is deterministic, as typically provided by \gls{pos}-based blockchains.

Synchronized smart contracts may also depend on other smart contracts.
Synchronizing only a single contract without considering dependencies could lead to unexpected behavior.
Therefore, all referenced contracts and libraries must be migrated and synchronized.
Previous work on smart contract forks proposed the analysis of logic and state during migration and recursively deploying dependencies~\cite{westerkamp2019portability}.
As contract addresses change during deployment, they are substituted for each reference.
Replacing addresses in the state results in a distinct state and thus Merkle root, leading to complex verification during deployment.
In case of synchronizations, the same mechanism could be applied.
As the contract's Merkle root is deemed valid after migration, the computed transition confirmation also utilizes a valid root and does not need to be modified to create valid results.
Therefore, the presented concept is applicable even if reference addresses were replaced during the initial migration.

SmartSync promotes read-only access for synchronized smart contracts and does not provide cross-chain write operations.
While other approaches target such write access~\cite{robinson2021general,nissl2021crosschain}, multiple steps including locking is required to maintain atomicity.
As a result, the process is time-intensive and may lead to additional overhead in case the timeout period is exceeded.
Therefore, our solution leaves the responsibility of invoking transactions across multiple blockchains to the external application.
Depending smart contracts hosted on another blockchain network are put in a desired state that is subsequently proven on the target blockchain.
Hereby, the resulting state becomes accessible through simple function calls.
Time consuming locking mechanisms are mitigated and instant function execution is enabled on the target blockchain.

\section{Conclusion}
In this paper, we presented SmartSync, a novel concept for synchronizing smart contracts across multiple blockchain networks.
The state of a remotely hosted smart contract is synchronized periodically in a verifiable manner.
No trust in executing entities is required, enabling all participants to perform state updates.
Furthermore, we introduced transition confirmations that prove correct smart contract state transitions on a secondary blockchain based on a multi proof.
As the entire smart contract is migrated from the source blockchain, the contract's state and logic become accessible on the target network.
Hereby, instant read-only function calls by depending smart contracts are rendered possible, facilitating use cases such as cross-chain registries or oracles.
Our open source implementation proves the concept's soundness and demonstrates timely updates of contract states. 

In the future, the presented solution will foster overarching architectures that include off-chain components as well as on-chain workflows executed on multiple blockchains.
Providing atomicity within such workflows depicts a potential task for future research.
Required state updates are initiated sequentially and become available after synchronization.
Thus, SmartSync constitutes a foundation for applications that utilize smart contracts across multiple blockchain networks.

\bibliographystyle{IEEEtran}
\bibliography{references}

\end{document}